\documentclass[aps,floats,amssymb,twocolumn,showpacs,superscriptaddress]{revtex4}
\usepackage{calc}
\usepackage{psfrag}
\usepackage{graphicx}
\begin{document}
\begin{abstract}
We discuss  t-J-U model on a honeycomb monolayer that has the same low-energy description of the kinetic term as graphene bilayer, and in particular study coexistence of antiferromagnetism and superconducting correlations that originate from Cooper pairs without phase coherence. We show that the model is relevant for the description of graphene bilayer and that the presence of the $d + i d$ superconducting correlations with antiferromagnetism can lead  to quadratic dependence in small magnetic fields of the gap of the effective monolayer consistent with the transport measurements of Velasco et al. on the graphene bilayer.
\end{abstract}
\title{On the coexistence of antiferromagnetism and $d + i d$ superconducting correlations in the graphene bilayer}
\author{M.V. Milovanovi\'c}
\affiliation{Scientific Computing Laboratory, Institute of Physics
Belgrade, University of Belgrade, Pregrevica 118, 11 080 Belgrade,
Serbia}
\author{S. Predin}
\affiliation{Department of Physics, University of Belgrade, P.O. Box 368, 11001
Belgrade, Serbia}
\maketitle
\section{Introduction} The interaction effects are important for the physics of graphene bilayer; recent experiments reveal gapped phase(s) in the undoped graphene bilayer which without interactions would represent a gapless system.
In a recent experiment, in Ref. \onlinecite{Lau}, with high quality samples, a completely insulating behavior was detected in transport measurements.  Theoretical investigations, mean field and renormalization group approaches \cite{Min,Nandkishore,Zhang,Vafek,Vafek_two,FZhang,Jung,Lemonik,ZhangMacD,Vafek_three,Scherer,
Lemonik2,Gorbar,Cvetkovic,ZhMin}, speak for a close competition of a few, mostly gapped, phases. One of the most prominent candidates for an explanation of the experiment in Ref. \onlinecite{Lau} is a layer  antiferromagnetic (LAF) state. The main reason for the existence of this state would be an on-site Coulomb repulsion, $U$; indeed as
pointed out in Ref. \onlinecite{Vafek_two}, a Hubbard model on a honeycomb bilayer lattice would lead to the LAF state, both in weak and strong $U$ limit. This may remind us of the behavior of the Hubbard model on the square lattice and the antiferromagnetic (AF) behavior due to nesting in the weak coupling limit. The Hubbard model on a square lattice is usually invoked as a model for cuprates in its strong coupling limit which forbids the double occupancy and leads to a ``perfect" AF behavior at half filling. On the other hand the estimate for $U$ is hard to know in the graphene bilayer and certainly depends on the computational scheme but it is expected to be stronger than both (inter and intra-layer) hoppings. Due to the smallness of the gap revealed in the experiment in Ref. \onlinecite{Lau} we will not consider the large $U$ limit (exclusion of double occupancy) when modeling graphene bilayer. But we will keep the on-site repulsion as a main
cause of the insulating behavior detected in the experiment. As expected from previous approaches this will lead to AF insulating behavior but seems not sufficient to describe all phenomena detected in the experiment.
An additional order parameter, besides the one that describes the antiferromagnetism is necessary for the complete explanation of the transport data of the experiment \cite{Lau,Vafek_three}.

In this work we will look for the additional order parameter that can coexist with antiferromagnetism in the graphene bilayer at half filling. We will argue that this is $d + i d$ (broken time reversal symmetry) - wave superconducting order parameter.
%On the level of the Hamiltonian description we will find particle-hole asymmetry and asymmetry under reversal of the %direction of perpendicular magnetic field, which we believe are essential for the explanation of the experimental data %of Ref.\onlinecite{Lau}.
This ($d + i d$) order parameter and its coexistence with antiferromagnetism was already found at finite (non-zero) dopings in a numerical (Grassman tensor product) approach to $t - J$ (large $U$) model on the honeycomb monolayer in Ref. \onlinecite{Gu}.  Due to the assumed moderate (not large) value of $U$ in our model of graphene bilayer the AF and $d + i d$ superconducting order parameter can coexist even at half-filling. Our model of the graphene bilayer can be described as a t-J-U model on an effective honeycomb lattice and, in the following, we will argue why this model is relevant for the description of graphene bilayer.
%The derivation of the model will not be rigorous, but will motivate the inclusion %of
%processes that are not emphasized in the weak coupling approaches in the %literature.

\section{Model and its motivation}
The kinetic part of the Hamiltonian that describes the garphene bilayer on two honeycomb lattices, which are Bernal stacked, is
\begin{eqnarray}
H_{0} &= & - t \sum_{\vec{n},\sigma} \sum_{\vec{\delta}}
(a_{1,\vec{n},\sigma}^{\dagger} b_{1, \vec{n} + \vec{\delta},\sigma}
+ a_{2,\vec{n},\sigma}^{\dagger} b_{2, \vec{n} -\vec{\delta},\sigma}
+ h.c.) \nonumber \\
&& + t_{\bot} \sum_{\vec{n},\sigma}
(a_{1,\vec{n},\sigma}^{\dagger} a_{2, \vec{n},\sigma} + h.c).\label{Hham}
\end{eqnarray}
The index $i = 1,2$ denotes the layer index. In Fig. 1 the relative
positions of two triangular sublattices, $A_1$ and $B_1$, for the
lattice 1, and $A_2$ and $B_2$, for the lattice 2 are illustrated.
In Eq.(\ref{Hham}) $t$ is the hopping energy between nearest
sites
in
each layer, and $t_{\bot}$ is the same energy for hopping between
the layers.
%The direct hopping between B1 and B2 sublattice is
%assumed to be zero. This assumption will lead to quadratic
%dispersion of electrons when  $t_{\bot} \gg t$ even for lowest
%momenta.
The on-site creation (annihilation) operators,
$a_{i,\vec{n},\sigma}^{\dagger} (a_{i, \vec{n},\sigma})$, are for
the electrons in the sublattice $A_i$ of the layer $i$ with spin
$\sigma = \uparrow, \downarrow$, and $b_{i,\vec{n},\sigma}^{\dagger}
(b_{i, \vec{n},\sigma})$ for the electrons in the sublattice $B_i$. $\vec{\delta}$'s are defined as $\delta_{1} = a
(0, 1/\sqrt{3})$, $\delta_{2} = a/2 (1, - 1/\sqrt{3})$, and
$\delta_{3} = a/2 (- 1, - 1/\sqrt{3})$, and $ a = \sqrt{3}\;
a_{cc}$, $a_{cc}$ is the distance between sites and $a$ is the
next to nearest neighbor distance.
\begin{figure}
\centering
\includegraphics[scale=0.3]{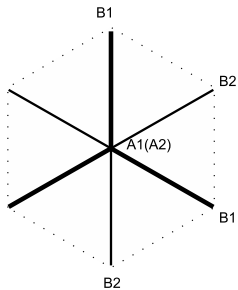}
\caption{A view of Bernal stacked honeycomb lattices 1 and 2 with
corresponding sublattice sites A1 and B1, and A2 and B2.}
\end{figure}

According to the results of renormalization group approaches \cite{Vafek,Vafek_two,Vafek_three,Cvetkovic,Lemonik,Lemonik2} the
antiferromagnetic (LAF) state and a nematic state are main instabilities that may arise due to interactions in the graphene bilayer. The nematic state is a result of an ordering in the particle-hole channel that can be described by a non-zero
expectation value of the hopping (bond) operator,
\begin{equation}
\sum_\sigma b_{1, \vec{n} + \vec{\delta}_1,\sigma}^{\dagger} b_{2, \vec{n} - \vec{\delta}_2,\sigma}, \label{bond}
\end{equation}
where $\vec{n} + \vec{\delta}_1$ and $\vec{n} - \vec{\delta}_2$ denote sites that are near neighbors on the honeycomb lattice that make sublattices $B1$ and $B2$, as shown in Fig. 1. On the same lattice the antiferromagnetic ordering can occur which describes the LAF state.

The nematic i.e. bond ordering as described in Ref. \onlinecite{Lemonik} can be thought as a $d_{x^2-y^2}$ CDW (or a $d + i d$ CDW as described in Ref. \onlinecite{Vafek} - see below Eq.(\ref{possibilities}) for the definition of the $d_{x^2-y^2}$ and $d + i d$ ordering) in the language of Ref. \onlinecite{nayak}. In the same reference this instability (``hidden order" on a square lattice) was proposed for cuprates. Just as in the case of the square lattice and cuprates, this ordering, in the graphene bilayer, can be thought as the result of short-range interactions and superexchange processes. These interactions and processes can also lead to the LAF state. The effective Heisenberg interaction on the $B1$ and $B2$ honeycomb, of neighboring sites $i$ and $j$ ($\vec{n} + \vec{\delta}$ and $\vec{n} - \vec{\delta}'$) can be rewritten in terms of the hopping operators (Eq.(\ref{bond})) as
\begin{equation}
 \vec{S}_i  \vec{S}_j = - \frac{1}{2} ( \sum_\sigma b_{1 i \sigma}^{\dagger} b_{2 j \sigma}) ( \sum_\sigma b_{2 j \sigma}^{\dagger} b_{1 i \sigma})- \frac{1}{4},
 \end{equation}
 and should be a part of an effective description of the graphene bilayer.
 According to Refs. \onlinecite{Vafek_two} and \onlinecite{Cvetkovic}, the only instability in the weak coupling limit of the Hubbard model on the graphene bilayer is antiferromagnetism. Thus, on phenomenological grounds, we will assume that the value of $J$ is an independent parameter in the effective description, favorable for both instabilities, nematic and antiferromagnetic.

 With respect to cuprates the on-site repulsion, $U$, is not that strong to preclude the double occupancy on the graphene bilayer. Therefore we will explicitly include this interaction in our model Hamiltonian, which can be described as
 \begin{equation}
H = H_{\text{J}} + H_{\text{H}} \label{JH},
\end{equation}
 where
 \begin{eqnarray}
 H_{\text{J}} = J \sum_{\vec{n}}\sum_{\vec{\delta_1},\vec{\delta_2}} \vec{S}_{b1} (\vec{n}+ \vec{\delta}_1)\;
\vec{S}_{b2} (\vec{n}- \vec{\delta}_2)\label{modelH},
\end{eqnarray}
and the summation is over the near neighbors, and
\begin{eqnarray}
H_{\text{H}} = - \sum_{k \in IBZ}\; t_{\text{eff}} \; \gamma_k^2 b_{1 k}^{\dagger}\; b_{2 k} + h.c.
 + \sum_{i = 1,2} \sum_{\vec{n}_i} U \hat{n}_{\uparrow \vec{n}_i} \hat{n}_{\downarrow \vec{n}_i}\label{HH}
\end{eqnarray}
with $\gamma_k = \sum_{\vec{\delta}} \exp\{i \vec{k} \vec{\delta}\}$,
  $\vec{n}_1 \equiv \vec{n} + \vec{\delta}_1$ and $\vec{n}_2 \equiv \vec{n} - \vec{\delta}_2$. The kinetic term (written above in the momentum space) in real space, on the effective honeycomb monolayer of $B1$ and $B2$ sites, describes near-neighbor and two times weaker third-neighbor hopping.  It can be recovered in the case of the noninteracting honeycomb bilayer by taking interlayer hopping to be large. In that case
$t_{\text{eff}} = \frac{t^2}{t_\bot}$. In the small momentum limit the kinetic term  in Eq.(\ref{HH}) becomes the one of the graphene bilayer, i.e. $t_{\text{eff}} \; \gamma_{k}^2 \rightarrow \frac{t^2}{t_\bot} (k_x \mp i k_y)^2$ near $K$ points: $K_{\pm} = \pm \frac{2\pi}{a}(\frac{2}{3},0)$ \cite{McFa,qhebg}. Because we look for the low energy properties we will keep the two-band extension with $\gamma_{k}^2$ throughout the Brillouin zone.

This model is similar to the $t-J-U$ model defined on  the square lattice and known from previous investigations. The $t-J-U$ model appeared also in the context of gossamer superconductivity \cite{VMC,Laugh,FCZhang,Yuan,Heiselberg}, the superconductivity that can exist even at half-filling.

\section{Mean field approach} In order to apply a mean field approach we can use the identity,
\begin{equation}
 \vec{S}_i  \vec{S}_j = - \frac{1}{2} (b_{i \uparrow}b_{j \downarrow} - b_{i \downarrow} b_{j \uparrow})(b_{j \downarrow}^\dagger b_{i \uparrow}^\dagger - b_{j \uparrow}^\dagger b_{i \downarrow}^\dagger)+ \frac{1}{4}.
 \end{equation}
 We will define the superconducting order parameter,
 \begin{equation}
 \Delta_{\vec{\delta}} = \langle b_{1 i \uparrow} b_{2 i+\vec{\delta} \downarrow} -  b_{2 i+\vec{\delta} \uparrow} b_{1 i \downarrow}\rangle,
 \end{equation}
 where $\vec{\delta}$ can be any of the near-neighbor vectors on the honeycomb lattice, and
 \begin{equation}
 m = \langle n_{i \uparrow}  - n_{i \downarrow} \rangle,
 \end{equation}
 the antiferromagnetic order parameter. In the following we will  use the following notation, $t \equiv t_{\text{eff}}$, and apply to the Hamiltonian in Eq.(\ref{JH}) the mean field ansatzes. We will generalize the derivation of Ref. \onlinecite{VTGM} to the case with the antiferromagnetic order parameter.
% \begin{eqnarray}
% &&H_{MF}  = - \sum_{k} t\; \gamma_k^2 b_{1 k}^{\dagger}\; b_{2 k} + h.c. %\nonumber \\
% &&\;- \frac{J}{2} \sum_{k} \sum_{\vec{\delta}} \Delta_{\vec{\delta}}
%  \exp\{i \vec{k} \vec{\delta}\} (b_{1 k \uparrow}^{\dagger} b_{2 -k %\downarrow}^{\dagger} - b_{1 k \downarrow}^{\dagger} b_{2 -k \uparrow}^{\dagger}) %+ h.c. \nonumber\\
% &&\; + \frac{1}{2} N J  \sum_{\vec{\delta}} |\Delta_{\vec{\delta}}|^2 %\nonumber\\
% &&\; + \frac{U}{2} \sum_{k} \sum_{i=1,2} (-1)^i m (\hat{n}_{i \vec{k} \uparrow} %- \hat{n}_{i \vec{k} \downarrow}) \nonumber\\
% &&\; + U \frac{m^2}{2} N,
% \end{eqnarray}

 If we use spinors,
 \begin{equation}
 \Psi_{k} = [ b_{1 k \uparrow}\;\; b_{2 k \uparrow}\;\; b_{1 -k \downarrow}^{\dagger}\;
 b_{2 -k \downarrow}^{\dagger}]^{T},
 \end{equation}
 we can write the mean field Hamiltonian as
 \begin{equation}
 H_{MF} = \Psi_{k}^{\dagger} {\cal H} \Psi_k + \frac{1}{2} N J  \sum_{\vec{\delta}} |\Delta_{\vec{\delta}}|^2 + U \frac{m^2}{2} N,
 \end{equation}
 where $N$ is the number of unit cells,
 \begin{equation}
{\cal H} = \left[\begin{array}{cccc}  \frac{U}{2} m & - t \gamma_k^2& 0 &  -\frac{J}{2} \Delta(k) \\
- t \gamma_k^{* 2} & - \frac{U}{2} m & -\frac{J}{2} \Delta(-k)& 0\\
0 & -\frac{J}{2} \Delta^{*}(-k)&  \frac{U}{2} m &  t \gamma_{-k}^{* 2}\\
-\frac{J}{2} \Delta^{*}(k) & 0 & t \gamma_{-k}^{2} & - \frac{U}{2} m
                                   \end{array} \right], \label{matrix}
\end{equation}
and $\Delta(k) = \sum_{\vec{\delta}} \Delta_{\vec{\delta}} \exp\{i \vec{k} \vec{\delta}\}$.
The symmetry analysis of the
order parameter on a honeycomb lattice, first done in Ref.
\cite{Black}, concluded that there are three possibilities,
\begin{eqnarray}
\Delta_{\vec{\delta}}: \begin{array}{ccc} \Delta \; (1,&1,&1) \\ \Delta \;(2,&-1,&-1)\\
\Delta \; (0,&1,&-1) \end{array}\label{possibilities}
\end{eqnarray}
that span the space of order parameter. The last two possibilities
belong to a two-dimensional subspace of irreducible representation
of ${\cal{S}}_3$, permutation group \cite{Polleti}.
%This means that
%any combination of these two order parameters is possible while
%keeping the symmetry intact.
 The $s$-wave, $\Delta_{\delta} = \Delta (1,1,1)$, has nodes at $K$ points because $\Delta_{k} = \Delta \gamma_{k}$. For $d_{x^2-y^2}$ wave, $\Delta_{\delta} = \Delta (2,-1,-1)$, near $K_{\pm}$ points we have $\Delta_{1}(K_\pm + k) = \Delta [3 \pm  \frac{\sqrt{3}}{2}(k_x \pm i k_y)]$, and for $d_{xy}$ wave, $\Delta_{\delta} = \Delta (0,1,-1)$, the expansions are
$\Delta_{2}(K_\pm + k) = \Delta i [\pm \sqrt{3} -  \frac{1}{2}(k_x \pm i k_y)]$. Therefore if the $d + i d$ combination, $\Delta_{1}(k) \pm i \sqrt{3} \Delta_{2}(k)$ is taken near one of the $K$ points the order parameter is a constant $(6 \Delta)$ and at the other $K$ point is linear in $k_x$ and $k_y$. Therefore, instead of having the coefficients of the same absolute magnitude with
$b_{1 k \uparrow}^{\dagger} b_{2 -k \downarrow}^{\dagger}$ and
$b_{2 k \uparrow}^{\dagger} b_{1 -k \downarrow}^{\dagger}$  (and $b_{2 -k \downarrow} b_{1 k \uparrow}$ and
$b_{1 -k \downarrow} b_{2 k \uparrow}$) for a fixed valley point $d + i d$ singles out one spin projection (up or down) to be associated with sites on layer 1 and the opposite one to be associated with sites on layer 2. Thus it favors pairing (Cooper pairs) in which layer index is associated with definite spin projection just as in an antiferromagnetically ordered state i.e. LAF state described above.

  $d + i d$ wave and $s$ wave can coexist, as rotationally symmetric states, with the LAF state although only for certain  values of $J$ and $U$ parameters. One can show that for the coexistence of the LAF state and $s$-wave $J \gg U$, which cannot be the case in the graphene bilayer. For $d + i d$ wave, on the other hand, one can find an interval for couplings, $J$ and $U$, for which the LAF state and $d + i d$-wave pairing can coexist.
Expanding the mean field equations to fourth order in the ratio of superconducting and antiferromagnetic order parameter and in the weak coupling limit i.e. $t > J,U$ we find $\frac{4}{3} - \frac{1}{4} Jw < \frac{J}{U} <
\frac{4}{3} + \frac{1}{4} J w$, with $w = \frac{1}{\sqrt{3} \pi t}$. We expect
that the interval will broaden when the short-range correlations (due to $U$) are properly taken into account that will renormalize (reduce) the effective value of $t$. This was worked out for square lattice in Ref. \onlinecite{Yuan} in a comprehensive (renormalized mean field) study of $t-J-U$ model, and a interval of couplings was identified for which antiferromagnetism and superconducting correlations can coexist at half-filling. Furthermore, in Ref. \onlinecite{VMC} variational Monte Carlo method was applied to the same system, and a finite value of the pairing amplitude $(\Delta)$ was found in the antiferromagnetic region (with no superconducting phase coherence). We expect a similar situation in our case.

As in the case of honeycomb monolayer in Ref.
\onlinecite{Black}, we can diagonalize the free part of  the above Hamiltonian and come to the following expressions for order parameters,
\begin{eqnarray}
C_{\vec{k}} &&= \sum_{\vec{\delta}}
\Delta_{\vec{\delta}} \cos\{\vec{k} \vec{\delta} - 2
\phi_{\vec{k}}\}\nonumber
\end{eqnarray}
and
\begin{eqnarray}
S_{\vec{k}} &&= \sum_{\vec{\delta}}
\Delta_{\vec{\delta}} \sin\{\vec{k} \vec{\delta} - 2
\phi_{\vec{k}}\},\nonumber
\end{eqnarray}
where $\phi_{\vec{k}} = arg(\gamma(k))$.
 Due to the expansion of $\Delta_k$ around $K$ points in the case of $d + i d$ wave we have
\begin{equation}
C_{K_\pm + k} \sim S_{K_\pm + k} \sim \frac{(k_x \pm i k_y)^2}{|k|^2},
\end{equation}
where the last sign is independent of $K$ points. Thus we recovered the basic signatures of $d + i d$ pairing; (a) the order parameter is  an eigenfunction
of orbital angular momentum with eigenvalue equal to two, and (b) due to the same sign (chirality) at both $K$ points this wave is time reversal symmetry breaking wave on the bilayer honeycomb lattice. Therefore in analyzing this wave we can keep the leading behavior in $\Delta_{1}(k)$ and $\Delta_{2}(k)$ as this effectively captures the basic phenomenology of $d + i d$-wave. Thus we  take
\begin{equation}
{\cal H}_{d-id} = \left[\begin{array}{cccc}  \frac{U}{2} m & - t \gamma_k^2& 0 &  - J 3 \Delta \\
- t \gamma_k^{* 2} & - \frac{U}{2} m & 0 & 0\\
0 & 0 &  \frac{U}{2} m &  t \gamma_{-k}^{* 2}\\
-J 3 \Delta  & 0 &  t \gamma_{-k}^{2} & - \frac{U}{2} m
                                   \end{array} \right],
                                   \label{h+}
\end{equation}
in the case of $\Delta(k) = \Delta_{1}(k) - i \sqrt{3} \Delta_{2}(k)$ combination, or
\begin{equation}
{\cal H}_{d+id} = \left[\begin{array}{cccc}  \frac{U}{2} m & - t \gamma_k^2& 0 & 0\\
- t \gamma_k^{* 2} & - \frac{U}{2} m & -J 3 \Delta  & 0\\
0 & -J 3 \Delta  &  \frac{U}{2} m &  t \gamma_{-k}^{* 2}\\
0 & 0 &  t \gamma_{-k}^{2} & - \frac{U}{2} m
                                   \end{array} \right],
                                   \label{h++}
\end{equation}
in the case when $\Delta(k) = \Delta_{1}(k) + i \sqrt{3} \Delta_{2}(k)$.
 %in the following mean field Hamiltonian
%\begin{equation}
% H_{MF} = \Psi_{k}^{\dagger} {\cal H}_{+} \Psi_k +  N J 6 \Delta^2 + U %\frac{m^2}{2} N
% \end{equation}
 In the same low-momentum limit $\gamma_{K_{\pm}+k} \approx \mp a \frac{\sqrt{3}}{2} (k_x \mp i k_y)$. We will use  redefinitions $ \frac{U}{2}m \equiv m$, $ J 3 \Delta \equiv \Delta,$ and $ ta^2 \frac{3}{4} \equiv t$ in the following.

We take $\Delta$ to be purely real and without the phase (U(1)) degree of freedom i.e. phase coherence \cite{RN} that would lead to supercurrents proportional to the gradient of this phase that would screen a magnetic field that may be present. We assume that supercurrents cannot develop in the antiferromagnetic, insulating background.
 %In other words we are assuming that a gauge transformation is done \cite{mfzt}, which takes the phase degrees of freedom to the kinetic part, and that after a mean field averaging, because of the incoherence of the phase, the only degree of freedom that is left is the amplitude of the order parameter. Therefore we will consider that Cooper pairs coexist with antiferromagnetism, but without the phase coherence i.e. fully developed superconductivity.

The Bogoliubov spectrum is the same irrespective whether we ask for energy eigenvalues in the case defined by Eq.(\ref{h+}) or  Eq.(\ref{h++}),
and with the introduced redefinitions the eigenvalues are
\begin{equation}
E = \pm \sqrt{m^2 + (\frac{\Delta}{2} \pm \sqrt{t^2 k^4 +(\frac{\Delta}{2})^2})^2}. \label{eigenvalues}
\end{equation}
Therefore the two different chirality states  of the $d$-wave are equally likely in the presence of the antiferromagnetic ordering.

\section{Presence of small magnetic field} In the presence of magnetic field, due to the minimal prescription, we may introduce a pair of creation and annihilation operators and express the resulting Hamiltonian matrix  around $K_+$ point as
 \begin{equation}
{\cal H}_{+ (d-id)B}^{o} =
 \left[\begin{array}{cccc}  m & -\omega_c (a^{\dagger})^2& 0 &  -\Delta \\
-\omega_c (a)^2 & -m & 0 & 0\\
0 & 0 &   m & \omega_c (a^{\dagger})^2\\
-\Delta  & 0 &  \omega_c (a)^2 & -m
                                   \end{array} \right].
                                   \label{h+B}
\end{equation}
Here we introduced $\omega_c = \frac{eB}{mc}$, where $B$ is the magnetic field and $m$ is the effective mass of the graphene bilayer, $ \frac{1}{m} = 2 t$.
The eigenvectors can be expressed as 4-spinor coefficients of eigenvectors
$\Psi_n$ of $a^\dagger a$ - operator, $a^\dagger a \Psi_n = n \Psi_n$, classified by integer eigenvalues $n$: $0, 1, 2, \ldots$. In the presence of small magnetic field we will look for the eigenstates in
the form
\begin{equation}
 \Psi_{n} = [ c_1 \; c_2 \; c_3 \;
 c_4 ]^{T} |n \rangle \;\; n = 0, 1, 2, \ldots \label{eq2}
 \end{equation}

The Nambu-Gorkov formalism with 4-spinors artificially doubles the degrees of freedom. This appears in spectra as doubling of energy levels $(\pm E)$. Thus when solving the ${\cal H}_{+ (d-id)B}$ we have to keep levels that are continuously related to energy levels with no superconducting instability $(\Delta \neq 0)$ and
are pertinent to the $2 \times 2$ upper, left block of the Hamiltonian matrix.

The Hamiltonian in Eq.(\ref{h+B}) we will consider under the approximation of a small magnetic field and rewrite it as
\begin{equation}
{\cal H}_{+ (d-id)B}^{o} = H_0 + V
\end{equation}
where
 \begin{equation}
H_0 =
 \left[\begin{array}{cccc}  m & 0 & 0 &  -\Delta \\
0 & -m & 0 & 0\\
0 & 0 &   m & 0\\
-\Delta  & 0 &  0 & -m
                                   \end{array} \right],
\end{equation}
and $V$ denotes the perturbation,
 \begin{equation}
V =
 \left[\begin{array}{cccc}  0 & -\omega_c (a^{\dagger})^2& 0 &  0 \\
-\omega_c (a)^2 & 0 & 0 & 0\\
0 & 0 &   0 & \omega_c (a^{\dagger})^2\\
0  & 0 &  \omega_c (a)^2 & 0
                                   \end{array} \right].
\end{equation}
Taking as solutions only values that are connected continuously in the limit $ \Delta \rightarrow 0$ to the upper $2 \times 2$ left part of $H_0$ we get for  the eigenvalues and eigenvectors of $H_0$:
\begin{eqnarray}
& E_1^n = - m  \;\;\; &\Psi_1 = [ 0, \; 1, \; 0, \;
 0 ]^{T} |n \rangle, \nonumber \\
 & E_2^n = \sqrt{m^2 + \Delta^2}   &\Psi_2 = c [ m + E, \; 0, \; 0, \;
 - \Delta ]^{T}  |n \rangle, \nonumber
 \end{eqnarray}
 where $E = \sqrt{m^2 + \Delta^2}$ and $ c = \frac{1}{\sqrt{2E(E+m)}}$.
 Considering the small magnetic field to second order as perturbation we get
 $E_1^n = - m - \frac{(n + 2)(n + 1)}{2} \frac{\omega^2_c}{E}$ and $E_2^n = E + \frac{n (n - 1)}{2} \frac{\omega^2_c}{E}$.
 Considering the same problem at
 $K'\equiv -K$ point
 %i.e.
 %\begin{equation}
%{\cal H}_{- (d-id)B} = H_0' + V'
%\end{equation}
% with
% \begin{equation}
%H_0' =
% \left[\begin{array}{cccc}  m & 0 & 0 &  0 \\
%0 & -m & -\Delta & 0\\
%0 & -\Delta &   m & 0\\
% 0 & 0 &  0 & -m
%                                   \end{array} \right],
%\end{equation}
%and
% \begin{equation}
%V' =
% \left[\begin{array}{cccc}  0 & -\omega_c (a)^2& 0 &  0 \\
%-\omega_c (a^{\dagger})^2 & 0 & 0 & 0\\
%0 & 0 &   0 & \omega_c (a)^2\\
%0  & 0 &  \omega_c (a^{\dagger})^2 & 0
%                                   \end{array} \right].
%\end{equation}
%We have similarly for $H_0'$ eigenvalues and eigenvectors,
%\begin{eqnarray}
%& E_1^n =  m  \;\;\; &\Psi_1 = [ 1, \; 0, \; 0, \;
% 0 ]^{T} |n \rangle, \nonumber \\
% & E_2^n = - \sqrt{m^2 + \Delta^2}   &\Psi_2 = c [ 0, \; E + m, \;\Delta,\; 0
%  ]^{T} |n \rangle .  \nonumber
% \end{eqnarray}
%Considering the small magnetic field to second order as perturbation
we get
 $\tilde{E}_1^n =  m + \frac{(n + 2)(n + 1)}{2} \frac{\omega^2_c}{E}$ and $\tilde{E}_2^n = - E - \frac{n (n - 1)}{2} \frac{\omega^2_c}{E}$.

 Thus, by analyzing the spectra of both $K$ points together, we can conclude that with the inclusion of small magnetic fields the gap changes from $E_g = 2 m$ value to $E_g = 2 m + 2 \frac{\omega^2_c}{E}$ in the presence of $d - i d$ correlations. Without the correlations or with $d + i d$ correlations the gap will not have the correction quadratic in small magnetic field, which direction is fixed in Eq.(\ref{h+B}). $d - i d$ correlations minimize the energy of the system by shifting also the energy levels closest to the Fermi  point. In the Appendix  we compare the energies  of the states with $d + i d$ and $d - i d$ correlations, and show that $d - i d$ are indeed of the lower energy.

  The energy minimization, when the direction of perpendicular magnetic field is opposite requires that the superconducting correlations are of $d + id$ kind. Thus the change in the direction of magnetic field is followed by the change in the chirality of the superconducting instability i.e. $ B \rightarrow - B$ followed by $ d - i d \rightarrow d + i d$ as one would expect from the superconducting instability that has orbital and therefore magnetic moment. This amounts to just switching of the previously found spectra
 between $K$ and $K'$ points. The gap is the same irrespective of the direction of the magnetic field although with the inclusion of superconducting correlations linear in $k$ (on the diagonal in the Eq.(\ref{h+}) and Eq.(\ref{h++})) leads to an asymmetry which may be related to  the asymmetry detected in the transport measurements
 of Ref. \onlinecite{Lau} with respect to the change in the direction of the external field.
 \section{Discussion and conclusion}

In the literature we find several proposals, Refs. \onlinecite{Kharitonov,Zhu,Roy},
for the explanation of the data of Ref. \onlinecite{Lau}.
See also Ref. \onlinecite{Bao} for further experimental investigations on the
same system and possible explanations based on anomalous quantum Hall physics.
 Ref. \onlinecite{Kharitonov} by Kharitonov introduces
an additional order parameter to the Neel order parameter, but the resulting gap dependence
does not have a minimum at B(magnetic field)=0 - compare Fig. 3
in Ref. \onlinecite{Kharitonov}, in contrast to what can be seen from the transport
measurements - compare Fig. 3 in Ref. \onlinecite{Lau}. Ref. \onlinecite{Zhu}, by Zhu, Aji,
and Varma, with the interesting proposal of taking into account
the full four band structure, still gives linear dependence on B
of the gap - compare with Fig. 6 in Ref. \onlinecite{Zhu} in contrast to the
quadratic dependence on small B as seen in the experiment. Ref. \onlinecite{Roy} by Roy does describe the quadratic dependence based on a mean
field treatment of spin magnetism, where a phenomenological
ferromagnetic interaction next to the Neel ordering among the spins of electrons is
introduced which existence (with a precise magnitude) is necessary
to obtain a correspondence to the experimental data.

Our approach is also mean field and phenomenological, though
clearly motivated microscopically by the physics of the $t-J-U$
model, as we introduce orbital magnetism of superconducting
correlations. This leads to the quadratic dependence of
the gap on small B as observed in the experiment. Thus we demonstrated a possibility that  the quadratic dependence on small magnetic field observed in the experiment of Ref. \onlinecite{Lau} may be due to the time reversal symmetry breaking $d$-wave superconducting correlations that coexist with antiferromagnetism.

%We showed that Heisenberg model and superexchange processes may be relevant for %the physics of the graphene bilayer. Depending on the strength of $U$, they may %lead also to the nematic state that competes with the antiferromagnetism as %revealed by renormalization group calculations \cite{Vafek,Lemonik2}, either as
%$d_{x^2-y^2}$ CDW as described in Ref. \onlinecite{Lemonik2}, or $d + i d$ CDW as %described in Ref. \onlinecite{Vafek}. Thus a mean field $t-J-U$ ansatz, based %also on the equality,
%\begin{equation}
% \vec{S}_i  \vec{S}_j = - \frac{1}{2} ( \sum_\sigma b_{i \sigma}^{\dagger} a_{j %\sigma}) ( \sum_\sigma a_{i \sigma}^{\dagger} b_{j \sigma})- \frac{1}{4},
% \end{equation}
%would lead to a more complete phase diagram.

The $d + i d$ wave superconductivity and antiferromagnetism at high dopings
of the graphene monolayer was studied in Refs. \onlinecite{Chubukov} and \onlinecite{RT}. It was shown \cite{RT} that both instabilities are connected with the existence of the on-site repulsion, U.

The last and important question we would like to discuss is how our proposal can explain the behavior  of the system gap in strong (and moderate) magnetic fields. In other words the question is how does the antiferromagnetic ground state with $d + i d$ superconducting correlations evolve in the many-body state of half-filled
zero-energy Landau level, which is eightfold degenerate due to flavor (spin and valley(layer)) and orbital ($n = 0,1$ Landau index) degrees of freedom. We expect
 a gradual formation of a QHFM (quantum Hall ferromagnet) \cite{Barlas} due to many-body correlations and the spontaneous ferromagnetic ordering of the spin degree of freedom. Thus we will fix the valley and orbital degree of freedom in the following and discuss how from two (spin up and spin down) Landau levels we can have effectively a single filled Landau level and ferromagnetic ordering. $d + i d$ wave Cooper pairs, described in the long-distance with the following Cooper pair wave function \cite{rg}
\begin{equation}
f(\vec{r}_{1 \uparrow} - \vec{r}_{2 \downarrow}) \sim \frac{\bar{z}_{1 \uparrow} - \bar{z}_{2 \downarrow}}{z_{1 \uparrow} - z_{2 \downarrow}},
\end{equation}
in the presence of the magnetic flux will be modified by flux (vortex) attachment due to the particles of opposite spin as in
\begin{equation}
f(\vec{r}_{1 \uparrow} - \vec{r}_{2 \downarrow}) \sim \frac{\bar{z}_{1 \uparrow} - \bar{z}_{2 \downarrow}}{z_{1 \uparrow} - z_{2 \downarrow}} \prod_i  (z_{1 \uparrow} - z_{i \downarrow})^2 \prod_j  (z_{2 \downarrow} - z_{j \uparrow})^2.
\end{equation}
This will lead to the following many-body state
\begin{equation}
Det(\frac{\bar{z}_{i \uparrow} - \bar{z}_{j \downarrow}}{z_{i \uparrow} - z_{j \downarrow}}) \prod_{i,j}  (z_{i \uparrow} - z_{j \downarrow})^2 = \prod_{i,j}  (z_{i \uparrow} - z_{j \downarrow}) \chi_2 \label{jain}
\end{equation}
where $Det$ denotes the determinant of antisymmetrized product of Cooper pair wave functions, and $\chi_2$ denotes the filled second Landau level wave function in the Jain notation. The
identity used in Eq.(\ref{jain}) was proved in Ref. \onlinecite{moran}. The topological properties of the wave function in Eq.(\ref{jain}) (or the low energy
properties of the system described with the wave function as discussed in Ref. \onlinecite{Read}) are equivalent to the Halperin (1,1,1) state or QHFM i.e. the following state
\begin{equation}
\prod_{i<j}  (z_{i \uparrow} - z_{j \uparrow})
\prod_{p<q}  (z_{p \downarrow} - z_{q \downarrow})
\prod_{l,m}  (z_{l \uparrow} - z_{m \downarrow}),
\end{equation}
for fixed valley and orbital index, and thus lead to the QHFM state with the effective filling factor $\nu_{\text{eff}} = 4$. It was shown in Ref. \onlinecite{Nand} that this state would lead to the gap with linear dependence on the (strong)
magnetic field as observed in Ref. \onlinecite{Lau}. Thus we described a possible route from antiferromagnetic state with $d + i d$ superconducting correlations to the spin QHFM state consistent with the experiment.

M.V.M. thanks M. Goerbig, D. Tanaskovi\'c, and J. Vu\v ci\'cevi\'c for previous collaboration.
The authors are supported by the Serbian Ministry of Education and Science under
project No. ON171017.

\appendix
\section*{Energy minimization}
To find whether $d - id$ or $d + id$  SC correlations coexist with antiferromagnetism in the presence of the magnetic field, which direction is
defined as in Eq.(\ref{h+B}), we should compare the two ground state energies,
\begin{eqnarray}
E^{d - id} = && \sum_{n=0}^{n'} [- m - (n + 1)(n + 2) \delta] + \nonumber \\
&& \sum_{n=0}^{n''} [- E - n (n - 1) \delta],
\end{eqnarray}
and
\begin{eqnarray}
E^{d + id} = && \sum_{n=0}^{\bar{n}'} [- E - (n + 1)(n + 2) \delta] + \nonumber \\
&& \sum_{n=0}^{\bar{n}''} [- m - n (n - 1) \delta],
\end{eqnarray}
where $ \delta = \frac{\omega^2_c}{2 E}$, and the bounds for the summations are
determined by the lower cut-off, $- E_c$, i.e. we have, in the $d - i d$ case,
\begin{equation}
(n' + 2)(n' + 1) = \frac{E_c - m}{\delta}
\end{equation}
and
\begin{equation}
  n'' (n'' - 1) = \frac{E_c - E}{\delta},
\end{equation}
and
\begin{equation}
(\bar{n}' + 2)(\bar{n}' + 1) = \frac{E_c - E}{\delta}
\end{equation}
and
\begin{equation}
  \bar{n}'' (\bar{n}'' - 1) = \frac{E_c - m}{\delta},
\end{equation}
in the $d + i d$ case.
After a few steps of simple algebra we get
\begin{equation}
 E^{d - id} - E^{d + id} = 2(m - E).
\end{equation}
Because $ E > m$, the energy minimization favors $d - i d$ SC correlations for the fixed direction of the magnetic field (Eq.(\ref{h+B})).
\end{document}